**Minibeam-pLATTICE: A novel proton LATTICE modality using minibeams**

Nimita Shinde, Weijie Zhang, Yuting Lin, and Hao Gao*

Department of Radiation Oncology, University of Kansas Medical Center, USA

* Corresponding author:

Hao Gao, Department of Radiation Oncology, University of Kansas Medical Center, USA.

Email address: hgao2@kumc.edu

**Ethical Statement:** This research was carried out under Human Subject Assurance Number 00003411 for University of Kansas in accordance with the principles embodied in the Declaration of Helsinki and in accordance with local statutory requirements. Consent was given for publication by the participants of this study.

**Abstract.**

**Background:** LATTICE, a form of spatially fractionated radiation therapy (SFRT) that delivers high-dose peaks and low-dose valleys within the target volume, has been clinically utilized for treating bulky tumors. However, its application to small-to-medium-sized target volumes remains challenging due to beam size limitations.

**Purpose:** To address the challenge of applying LATTICE radiation therapy to small-to-medium-sized targets, this work proposes a novel proton LATTICE (pLATTICE) modality using minibeams, namely minibeam-pLATTICE, that can extend the LATTICE approach for small-to-medium target volumes.

**Methods:** Three minibeam-pLATTICE methods are introduced. (1) M0: a fixed minibeam aperture orientation (e.g., 0º) for all beam angles; (2) M1: alternated minibeam aperture orientations (e.g., between 0º and 90º), for consecutive beam angles; (3) M2: multiple minibeam aperture orientations (e.g., 0º and 90º) for each beam angle. The purpose of M1 or M2 is to correct anisotropic dose distribution at lattice peaks due to the planar spatial modulation of minibeams. For each minibeam-pLATTICE method, an optimization problem is formulated to optimize dose uniformity in target peaks and valleys, as well as dose-volume-histogram-based objectives. This optimization problem is solved using iterative convex relaxation and alternating direction method of multipliers (ADMM).

**Results:** Three minibeam-pLATTICE methods are validated to demonstrate the feasibility of minibeam-pLATTICE for two clinical head-and-neck (HN), one abdominal and one brain case. The advantages of this modality over conventional beam (CONV) pLATTICE are evaluated by comparing peak-to-valley dose ratio (PVDR) and dose delivered to organs at risk (OAR). All three minibeam-pLATTICE modalities achieved improved plan quality compared to CONV, with M2 yielding the best results. For instance, in one HN case, the following improvements were observed: PVDR increased to 3.73 (M2), compared to 3.27 (CONV), 3.72 (M0), and 3.49 (M1), while the mean dose to the mandible was reduced to 0.18 Gy (M2), compared to 0.33 Gy (CONV), 0.17 Gy (M0), and 0.14 Gy (M1).

**Conclusions:** A novel minibeam-pLATTICE modality is proposed that can generate lattice dose patterns for small-to-medium target volumes, which are not achievable with conventional pLATTICE due to beam size limitations. Peak dose anisotropy due to 1D planar minibeam apertures is corrected through inverse treatment planning with alternating or multiple minibeam apertures per beam angle.

## 1. Introduction

LATTICE radiation therapy [1], a form of spatially fractionated radiation therapy (SFRT) [2], divides the target volume into regions of high-dose (peak) and low-dose (valley) areas, delivering the dose according to the prescribed values for each. LATTICE can be viewed as a 3D extension of GRID [3]. LATTICE has been routinely used for treating bulky tumors with large target volumes [4]. However, applying LATTICE to small-to-medium target volumes remains challenging. This difficulty arises from the challenge of creating multiple well-separated peak regions, particularly when the target volume is small or located near critical organs at risk (OAR). A specific example is head-and-neck (HN) case, where the tumor is often close to the brainstem. In our experiments, it was observed that the conventional proton LATTICE radiation therapy delivered a maximum dose of 16.57 Gy to the brainstem.

Proton minibeam radiation therapy (pMBRT) [5,6] independently delivers highly heterogeneous dose distributions using sub-millimeter beamlets generated via multi-slit collimators (MSC). These beamlets are spaced a few millimeters apart and are known to induce distinct radiobiological responses compared to conventional beam delivery. While pMBRT can generate spatially fractionated dose distributions even in small target volumes, traditional implementations are constrained to planar or linear geometries, limiting their ability to shape dose distributions in three dimensions.

This limitation motivates the integration of LATTICE principles with minibeam delivery to create spatially fractionated dose distributions that are not only heterogeneous but also volumetrically optimized, creating multiple well-separated peak regions within small-to-medium target volumes. While recent advancements have been made in conventional beam based proton LATTICE (pLATTICE) [7-10], minibeam based pLATTICE has not yet been explored.

This work proposes a novel minibeam-pLATTICE modality that enables the delivery of LATTICE dose patterns to small-to-medium target volumes by leveraging the fine spatial resolution of proton minibeams. This approach differs from earlier lattice studies that used fixed brass grid collimators [11] by allowing flexible, beam-angle-specific multi-slit collimation strategies. The proposed method incorporates inverse planning to optimize peak and valley dose conformity while addressing anisotropy caused by single-orientation collimators. Specifically, multiple models are introduced that use alternating or simultaneous minibeam orientations per beam angle to overcome the limitations of planar dose modulation. As demonstrated in Section 3, the proposed minibeam-pLATTICE model reduced the maximum brainstem dose from 16.57 Gy (in the conventional pLATTICE case) to 5.8 Gy.

## 2. Methods

### *2.1. Optimization formulation for minibeam-pLATTICE*

The inverse treatment planning problem for minibeam-pLATTICE can be formulated as

$$\min_{x} \; f(d)$$
$$\text{s.t.} \;\; x \in \{0\} \cup [G, +\infty\},$$
$$\text{d} \; = \; \text{Ax}. \tag{1}$$

In Eq. (1), the decision variable $x$ represents the spot intensity vector, $A$ represents the dose influence matrix, and d is the dose distribution. The first constraint in Eq. (1) is a minimum-monitor-unit (MMU) constraint [12, 13] that ensures plan deliverability by constraining the smallest non-zero value of $x$ to be $G$. The second constraint in Eq. (1) defines the dose distribution.

The objective function, $f(d)$, is defined as

$$f(d) = \frac{w_p}{n_{peak}} ||d_{peak} - b_{peak}||_2^2 + \frac{w_v}{n_{valley}} ||d_{valley} - b_{valley}||_2^2 + \sum_{i=1}^{N_1} \frac{\text{w}_{1i}}{n_i} ||d_{\Omega_{1i}} - b_{1i}||_2^2$$
$$+ \sum_{i=1}^{\text{N}_2} \frac{\text{w}_{2i}}{n_i} ||\text{d}_{\Omega_{2i}} - \text{b}_{2i}||_2^2 + \sum_{i=1}^{\text{N}_3} \frac{w_{3i}}{n_i} ||\text{d}_{\Omega_{3i}} - \text{b}_{3i}||_2^2 + \sum_{i=1}^{\text{N}_4} \frac{w_{4i}}{n_i} ||\text{d}_{\Omega_{4i}} - \text{b}_{4i}||_2^2.$$

The objective function in Eq. (1) defines the least square error between the actual doses and the dose constraints. We now describe each term in $f(d)$ briefly.

- In the first term of $f(d)$, $b_{peak}$ and $d_{peak}$ are the prescribed dose and actual dose delivered to peak region in the target volume respectively. Thus, the first term defines the least square error between the prescribed and actual peak dose. Similarly, the second term of $f(d)$ is the least square error between the actual valley dose ($d_{valley}$) and prescribed valley dose ($b_{valley}$). Note that, $n_{peak}$ and $n_{valley}$ are the number of voxels in the peak and valley regions of the target volume respectively.
- The third term in $f(d)$ enforces the dose volume histogram (DVH)-min constraint [14, 15] for the peak ($i = 1$) and valley ($i = 2$) regions of the target. The DVH-min constraint ensures that at least $p$% of the total voxels in each region receive a dose larger than the minimum dose $b_{1i}$. To define the DVH-min constraint, first define the active index set $\Omega_{1i}$ as $\Omega_{1i} = \{j | j \leq p \times n_i\}$ if $d'_{p \times n_i} \leq b_{1i}$, where $d'$ is the dose distribution $d$ sorted in descending order and $n_i$ is the number of voxels in the peak ($i = 1$) and valley ($i = 2$) regions. The set $\Omega_{1i}$ consists of indices of peak and valley regions which violate the DVH-min constraint. The term, $||d_{\Omega_{1i}} - b_{1i}||_2^2$, then defines the least square error between the actual dose delivered and minimum dose $b_{1i}$, for the indices in the active index set $\Omega_{1i}$.
- The fourth term describes $N_2$ DVH-max constraints [14, 15] for OAR. For any OAR $i$, the DVH-max constraint ensures that at most $p$% of the total voxels in OAR $i$ receive a dose larger than $b_{2i}$. The active index set $\Omega_{2i}$ is defined such that it contains indices of voxels in OAR $i$ that violate the DVH-max constraint. Mathematically, $\Omega_{2i} = \{j | j \geq p \times n_i\}$ if $d'_{p \times n_i} \geq b_{2i}$, where $d'$ is the dose distribution $d$ sorted in descending order and $n_i$ is the number of voxels in OAR $i$. Thus, the fourth term in $f(d)$ defines the error between actual dose and maximum allowed dose $b_{2i}$, for the voxel indices of OAR $i$ that violate the DVH-max constraint.

- The fifth term in $f(d)$ defines the least square error for OAR voxels that violates the D-max (dose-max) constraint. For any OAR $i$, the D-max constraint ensures that all the voxels in OAR $i$ receive dose less than or equal to $b_{3i}$. If there exist voxels in OAR $i$ that violate the constraint, then the active index set $\Omega_{3i} = \{j \in [n_i] | d_j \geq b_{3i}\}$ is non-empty. The term $||d_{\Omega_{3i}} - b_{3i}||_2^2$ defines the least square error for the voxels in OAR $i$ that violate the D-max constraint.
- The last term in $f(d)$ defines the least square error for OAR that violate the D-mean (dose-mean) constraint. For any OAR $i$, the D-mean constraint ensures that the mean dose delivered to all voxels in OAR $i$ is less than or equal to $b_{4i}$. If this constraint is satisfied, the active index set $\Omega_{4i}$ is empty. However, if the constraint is violated, then $\Omega_{4i} = [n_i]$, i.e., the active index set consists of all voxels in OAR $i$.

*2.2 Optimization methods for minibeam-pLATTICE*

This work introduces three minibeam-pLATTICE models (M0, M1 and M2), distinguished by the orientation of multi-slit collimators (MSC) relative to proton minibeams at different beam angles. Due to the highly directional nature of minibeams, the delivered dose exhibits anisotropic distribution within the target. To address this issue, this work explores using collimators with varying orientations to modulate the dose distribution, achieving a more spherical 3D peak dose pattern with spherical-shaped peak regions.

- **M0 (Baseline Model):** Utilizes a fixed horizontal collimator orientation (0º relative to the beam source) for all beam angles.

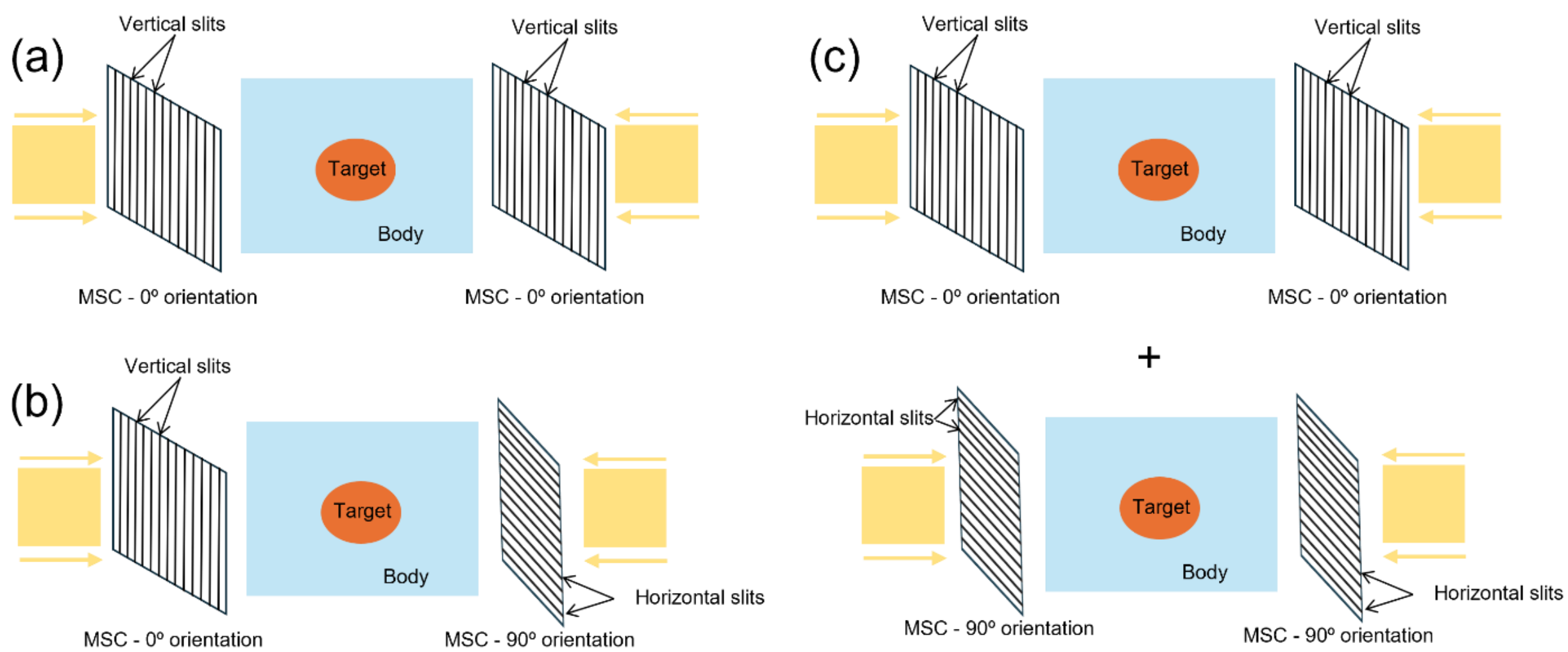

Figure 1: Three minibeam-pLATTICE models. (a) **M0** model with 0º MSC orientation for all beam angles, (b) **M1** model with alternating 0º and 90º MSC orientations, (c) **M2** model with both 0º and 90º MSC orientations at each angle.

- **M1 (Alternating Orientation Model):** Alternates between two orthogonal collimator orientations at different beam angles. In this study, two orthogonal orientations of 0º and 90º with respect to the beam sources are used.
- **M2 (Multi-Collimator Model):** Employs multiple collimators at each beam angle. In this study, both 0º and 90º orientations are used simultaneously at each beam angle. This results in two distinct minibeam apertures per beam angle, helping to counteract anisotropic dose distribution.

*2.3 Optimization algorithms for minibeam-pLATTICE*

To solve Eq. (1), auxiliary variable for the MMU constraint is introduced, and Eq. (1) is re-written as

$$\begin{aligned} &\min_{x}\ f(d) \\ &s.t.\ z \in \{0\} \cup [G, +\infty\}, \\ &z = x, \\ &\mathrm{d} = \mathrm{Ax}. \end{aligned} \tag{2}$$

Eq. (2) can be solved via iterative convex relaxation (ICR) method [16, 17] and alternating direction method of multipliers (ADMM) [18, 19]. This technique has been used to solve various inverse optimization problems in radiation therapy [20-29]. The iterative method described here involves updating the active index sets for the terms in $f(d)$, followed by updating each decision variable in Eq. (2) sequentially while keeping other variables fixed. To use the ADMM method, define the augmented Lagrangian for Eq. (2) as

$$\begin{aligned} &\min_{x,z}\ f(\mathrm{Ax}) + \frac{\mu_1}{2}||z - x + \lambda_1||_2^2 \\ &s.t.\ z \in \{0\} \cup [G, +\infty\}. \end{aligned} \tag{3}$$

Algorithm 1 is then used to solve the augmented Lagrangian formulation.

---

**Algorithm 1: Optimization method for solving Eq. (3)**

---

1. **Input:** Choose parameters $\mu_1, w_p, w_v, w_1, w_2, w_3, w_4$
2. Initialization: Randomly initialize $x$. Choose number of iterations $T$.
3. Set $\lambda_1 = z = x$.
4. For $t = 1, \ldots, T$
    a. Find active index sets $\Omega_1, \Omega_{2i}, \Omega_{3i}, \Omega_{4i}$ as described in Section 2.1.
    b. Update primal variables $x, z$ one at a time by fixing other variables and solving the resulting minimization problem.
    c. Update dual variable as $\lambda_1 = \lambda_1 + z - x$.
5. **Output:** $x$

---

In Step 4b of Algorithm 1, $x$ is updated by fixing the variable $z$. The resulting minimization problem is unconstrained in $x$. Thus, the optimal value of the decision variable $x$ is obtained by taking first-order derivative of the objective function and solving the linear system of equations. Similarly, to update $z$ in Step 4b, $x$ is fixed. The minimization problem in $z$ has the following closed form solution

$$z = \begin{cases} max\ (G, x - \lambda_1), & if\ x - \lambda_1 \geq G/2 \\ 0, & otherwise. \end{cases}$$

*2.4 Materials*

The performance of the conventional pLATTICE (CONV) model is compared with the three proposed minibeam-pLATTICE models, namely, M0, M1 and M2, across two head-and-neck (HN) cases, one abdominal case, and one brain case. Each test case is selected to capture anatomical variation and to test plan robustness under different geometric constraints, including highly concave targets and proximity to sensitive OAR such as the brainstem and kidney. Particularly, the HN and brain cases represent target volumes located in anatomically complex regions with nearby critical OAR, such as the brainstem and oral cavity, making them suitable candidates for evaluating spatially fractionated dose delivery techniques.

The prescribed peak and valley doses for each case are as follows: HN01 (2 Gy and 10 Gy), HN02 (2.12 Gy and 10.6 Gy), abdominal (1.8 Gy and 9 Gy) and brain (1.8 Gy and 9 Gy). The beam angles used are (0º, 45º, 90º, 135º, 180º, 225º, 270º, 315º) for HN01, (45º, 135º, 225º, 315º) for HN02 and brain cases, and (0°, 120°, 240°) for abdominal case. While conventional implementations often employ four gantry angles for HN cases, eight angles were used in HN01 to maintain consistency with the minibeam models, where additional angles are used to improve directional coverage, and enhance peak shaping.

In the minibeam-pLATTICE models, the number of spherical peak regions defined in the target volume are: 13 (HN01), 14 (HN02), 10 (abdominal), and 8 (brain), each with a diameter of 1.5 mm. In contrast, the conventional pLATTICE model includes 2 spherical peaks in each case, with 10 mm diameter for HN and abdominal cases, and 5 mm diameter for the brain case, constrained by tumor size. These peaks were manually placed within the target to maximize spatial separation and avoid overlap with adjacent OAR.

To ensure a consistent and fair comparison, both the conventional and minibeam-pLATTICE plans were optimized using the same iterative convex relaxation and ADMM framework. The only differences between models lie in the lattice geometries, collimation schemes, and beam configurations. Conventional pLATTICE plans used standard spot-scanning geometries without minibeam modulation, while the minibeam-pLATTICE models incorporated MSC with a 7 mm center-to-center slit distance and 0.4 mm slit width. The collimator orientations for M0, M1, and M2 follow the configurations described in Section 2.2. Dose influence matrices for each beam angle and source are generated using MatRad [30], with a spot width of 0.4 mm on a 1×1×1 $mm^3$ dose grid.

All plans are normalized to ensure that 95% of the valley region receives at least 100% of the prescribed valley dose. Plan quality is evaluated based on mean peak and valley doses, mean doses delivered to OAR, and peak-to-valley dose ratio (PVDR). The PVDR is computed as: PVDR = D10/D80 [31, 32], where D10 and D80 are the doses delivered to at least 10% and 80% of the entire target volume respectively.

Additionally, total spot weights are reported in giga-protons (Gp) as the sum of all spot intensities across all beamlets and beam angles. Total treatment session time is reported as the sum of (i) beam-on time, (ii) gantry setup time, and (iii) energy switching time. For the M1 and M2 models, an additional 30 seconds is included to account for collimator rotation and adjustment. To calculate the times, a proton dose rate of 400 giga-protons per second [33] is assumed, along with an estimated gantry rotation and setup time of 30 seconds per angle [33], and an energy switching time of 1 second per energy layer [34, 35]. For the M2 model, two collimator orientations are applied at each angle, resulting in two times the number of unique delivery configurations compared to other models. A 30-second gantry setup time is applied per configuration in this case.

## 3. Results

### *3.1 Dosimetric comparison: conventional pLATTICE vs. minibeam-pLATTICE*

Across all four clinical test cases, the minibeam-pLATTICE models (M0, M1, M2) consistently outperform the conventional pLATTICE (CONV) model in terms of objective function values, and peak-valley dose conformity. As shown in Tables 1-4, objective function values are significantly lower for minibeam plans, indicating improved conformity to DVH constraints (see Supplementary Material Section A for objective DVH terms). For example, in HN01, the objective function drops from 490.74 (CONV) to 85.16 (M2); in the abdominal case, it decreases from 55.84 (CONV) to 27.49 (M2). Minibeam models also provide better conformity to prescribed peak and valley doses. In all cases, the mean peak and valley doses for M0, M1, and M2 models are closer to prescription values compared to CONV. Furthermore, minibeam-pLATTICE plans consistently improve PVDR. For instance, PVDR increases from 3.27 (CONV) to 3.73 (M2) in HN02, from 4.13 to 4.94 in the abdominal case, and from 3.47 to 3.94 in the brain case.

OAR sparing is also enhanced in many scenarios. In HN01, the maximum brain dose drops significantly from 16.57 Gy (CONV) to 5.80 Gy (M2). In the abdominal case, M2 reduces spinal cord maximum dose from 3.05 Gy (CONV) to 2.30 Gy. While some OAR exhibit an increase in maximum dose delivered by M2 model (due to the highly spatially fractionated dose distribution inherent to minibeam delivery), mean doses to OAR and body generally remain comparable to or lower than the doses delivered in the conventional plan. For example, in HN01, mean body dose is reduced from 0.19 Gy (CONV) to 0.14 Gy (M2), however, the maximum dose delivered to body increases from 11.68 Gy (CONV) to 12.93 Gy (M2). Finally, the smaller objective function values for the minibeam-pLATTICE models demonstrate that the proposed models improve the adherence to OAR DVH constraints compared to the conventional model. These findings demonstrate that minibeam-pLATTICE, particularly the M2 model, offers enhanced target conformity and OAR sparing in most anatomical regions, reinforcing its dosimetric advantages over conventional pLATTICE approaches.

### *3.2 Comparison of different minibeam-pLATTICE methods*

Within the minibeam-pLATTICE modality, three models are proposed as described in Section 2.2. Among these models, M2 consistently offers the best balance between PVDR improvement and OAR sparing across all test cases. M2 achieves the lowest objective function values in three of four cases. For example, objective function value achieved by M2 is 27.49 in the abdominal case, achieving at least 41.97% improvement compared to other minibeam-pLATTICE models, indicating superior overall plan quality.

Additionally, the highest PVDR values are consistently obtained with M2. In the abdominal case, M2 achieves the highest PVDR of 4.94, slightly outperforming M0 (4.85) and M1 (4.86). M2 also maintains the lowest maximum doses to large bowel (8.97 Gy), spinal cord (2.30 Gy) and left kidney (10.74 Gy). Similarly, in HN02 case, the models achieve the PVDR of: M0 = 2.05, M1 = 2.31, and M2 = 2.55. In HN01, M2 further reduces the maximum dose to the brain and brainstem relative to the other models. In the brain case, M2 lowers the maximum dose to the brain to 10.26 Gy, compared to 15.65 Gy with M0, while still maintaining adequate target coverage despite slightly higher brain exposure than the CONV model.

All three minibeam models exhibit comparable performance in terms of mean doses to the body and other OAR, with M2 achieving an improved performance for most metrics in each case. The superior plan quality achieved with M2 is attributed to its use of multiple collimator orientations per gantry angle, enabling finer spatial modulation. Although M2 requires a greater number of delivery configurations, resulting in longer treatment times (e.g., 34.19 min for HN01), the dosimetric benefits justify the additional delivery complexity. The total spot weights across M0, M1, and M2 also shows that more sophisticated collimator strategies (M1 and M2) slightly increase the total spot count, but not disproportionately relative to the improvement in plan quality.

### *3.3 Treatment efficiency: total spot weights and session time*

The total spot weight (expressed in giga-protons) and the corresponding beam-on and session times are reported in Tables 1-4. As expected, the minibeam-pLATTICE models incur a significantly higher total MU due to the large number of narrowly collimated spots used to sculpt spatially fractionated dose distributions. For example, in HN01 case, total MU increases from 4.50e+4 (CONV) to 4.98e+5 (M2), and in HN02 case, from 4.76e+4 (CONV) to 6.21e+5 (M2). This results in longer beam-on times and overall session durations. For example, the total treatment time for M2 in HN01 is 34.19 minutes, compared to 8.54 minutes for CONV. These increases are attributed to both beam modulation complexity and hardware constraints such as energy layer switching and collimator repositioning. While this represents a trade-off in treatment efficiency, it reflects the improved dosimetric quality and spatial modulation achieved by the minibeam-pLATTICE approach.

Table 1: Comparison of conventional pLATTICE (CONV) and minibeam-pLATTICE (M0, M1, M2) for HN01 case. The prescribed peak and valley doses are 10 Gy and 2 Gy respectively.

| **Structure** | **Quantity** | **CONV** | **M0** | **M1** | **M2** |
|---|---|---|---|---|---|
| | Obj fn val | 490.74 | 95.11 | 80.62 | 85.16 |
| CTV | Mean $D_{peak}$ (Gy) | 15.14 | 11.69 | 11.48 | 11.64 |
| | Mean $D_{valley}$ (Gy) | 3.20 | 2.66 | 2.58 | 2.60 |
| | PVDR | 4.73 | 4.39 | 4.45 | 4.47 |
| Body | $D_{mean}$ (Gy) | 0.19 | 0.14 | 0.13 | 0.14 |
| | $D_{max}$ (Gy) | 11.68 | 17.64 | 16.95 | 12.93 |
| Brainstem | $D_{mean}$ (Gy) | 0.51 | 0.56 | 0.56 | 0.53 |
| | $D_{max}$ (Gy) | 3.93 | 7.20 | 8.10 | 6.52 |
| Brain | $D_{mean}$ (Gy) | 0.028 | 0.035 | 0.032 | 0.032 |
| | $D_{max}$ (Gy) | 16.57 | 6.54 | 7.04 | 5.80 |
| | Total spot wts (Gp) | 4.50e+4 | 5.05e+5 | 4.88e+5 | 4.98e+5 |
| | Beam-on time (secs) | 112.57 | 1264.22 | 1220.81 | 1247.43 |
| | Total time (mins) | 8.54 | 27.53 | 27.33 | 34.19 |

Table 2: Comparison of conventional pLATTICE (CONV) and minibeam-pLATTICE (M0, M1, M2) for HN02 case. The prescribed peak and valley doses are 10.6 Gy and 2.12 Gy respectively.

| **Structure** | **Quantity** | **CONV** | **M0** | **M1** | **M2** |
|---|---|---|---|---|---|
| | Obj fn val | 149.59 | 85.30 | 108.99 | 84.26 |
| CTV | Mean $D_{peak}$ (Gy) | 12.31 | 11.72 | 11.91 | 11.70 |
| | Mean $D_{valley}$ (Gy) | 3.76 | 3.15 | 3.41 | 3.14 |
| | PVDR | 3.27 | 3.72 | 3.49 | 3.73 |
| Body | $D_{mean}$ (Gy) | 0.036 | 0.032 | 0.033 | 0.032 |
| | $D_{max}$ (Gy) | 9.95 | 12.54 | 17.50 | 11.63 |
| Larynx | $D_{mean}$ (Gy) | 0.24 | 0.28 | 0.28 | 0.27 |
| | $D_{max}$ (Gy) | 2.27 | 2.60 | 2.53 | 2.52 |
| Mandible | $D_{mean}$ (Gy) | 0.33 | 0.17 | 0.14 | 0.18 |
| | $D_{max}$ (Gy) | 4.11 | 12.13 | 14.72 | 11.76 |
| Oral cavity | $D_{mean}$ (Gy) | 0.22 | 0.17 | 0.16 | 0.17 |
| | $D_{max}$ (Gy) | 2.10 | 2.52 | 2.88 | 2.84 |
| | Total spot wts (Gp) | 4.76e+4 | 6.04e+5 | 6.29e+5 | 6.21e+5 |
| | Beam-on time (secs) | 119.07 | 1512.13 | 1573.46 | 1553.72 |
| | Total time (mins) | 6.18 | 29.26 | 30.80 | 34.71 |

Table 3: Comparison of conventional pLATTICE (CONV) and minibeam-pLATTICE (M0, M1, M2) for pancreatic case. The prescribed peak and valley doses are 9 Gy and 1.8 Gy respectively.

| **Structure** | **Quantity** | **CONV** | **M0** | **M1** | **M2** |
|---|---|---|---|---|---|
| | Obj fn val | 55.84 | 47.06 | 39.03 | 27.49 |
| CTV | Mean $D_{peak}$ (Gy) | 10.11 | 10.24 | 9.99 | 9.68 |
| | Mean $D_{valley}$ (Gy) | 2.44 | 2.11 | 2.05 | 1.95 |
| | PVDR | 4.13 | 4.85 | 4.86 | 4.94 |
| Body | $D_{mean}$ (Gy) | 0.12 | 0.10 | 0.10 | 0.10 |
| | $D_{max}$ (Gy) | 11.55 | 20.52 | 18.40 | 13.49 |
| Large bowel | $D_{mean}$ (Gy) | 0.42 | 0.24 | 0.24 | 0.24 |
| | $D_{max}$ (Gy) | 4.38 | 11.45 | 11.06 | 8.97 |
| Spinal cord | $D_{mean}$ (Gy) | 0.42 | 0.39 | 0.38 | 0.36 |
| | $D_{max}$ (Gy) | 3.05 | 3.85 | 3.32 | 2.30 |
| Left Kidney | $D_{mean}$ (Gy) | 0.37 | 0.34 | 0.34 | 0.32 |
| | $D_{max}$ (Gy) | 5.90 | 11.30 | 11.21 | 10.74 |
| | Total spot wts (Gp) | 7.28e+4 | 8.69e+5 | 8.42e+5 | 8.16e+5 |
| | Beam-on time (secs) | 182.04 | 2174.81 | 2105.34 | 2040.83 |
| | Total time (mins) | 6.18 | 38.79 | 38.10 | 39.54 |

Table 4: Comparison of conventional pLATTICE (CONV) and minibeam-pLATTICE (M0, M1, M2) for HN01 case. The prescribed peak and valley doses are 9 Gy and 4.5 Gy respectively.

| **Structure** | **Quantity** | **CONV** | **M0** | **M1** | **M2** |
|---|---|---|---|---|---|
| | Obj fn val | 92.34 | 84.97 | 58.81 | 54.29 |
| CTV | Mean $D_{peak}$ (Gy) | 10.32 | 10.29 | 9.91 | 9.85 |
| | Mean $D_{valley}$ (Gy) | 2.97 | 2.74 | 2.52 | 2.50 |
| | PVDR | 3.47 | 3.75 | 3.93 | 3.94 |
| Body | $D_{mean}$ (Gy) | 0.05 | 0.05 | 0.05 | 0.05 |
| | $D_{max}$ (Gy) | 12.83 | 18.78 | 14.87 | 11.08 |
| Brain | $D_{mean}$ (Gy) | 0.12 | 0.20 | 0.18 | 0.18 |
| | $D_{max}$ (Gy) | 8.34 | 15.65 | 14.54 | 10.26 |
| Brainstem | $D_{mean}$ (Gy) | 0.07 | 0.08 | 0.08 | 0.09 |
| | $D_{max}$ (Gy) | 1.92 | 4.45 | 1.52 | 1.37 |
| | Total spot wts (Gp) | 2.33e+4 | 3.16e+5 | 3.04e+5 | 3.08e+5 |
| | Beam-on time (secs) | 58.23 | 792.04 | 760.65 | 771.17 |
| | Total time (mins) | 4.30 | 16.18 | 16.12 | 19.23 |

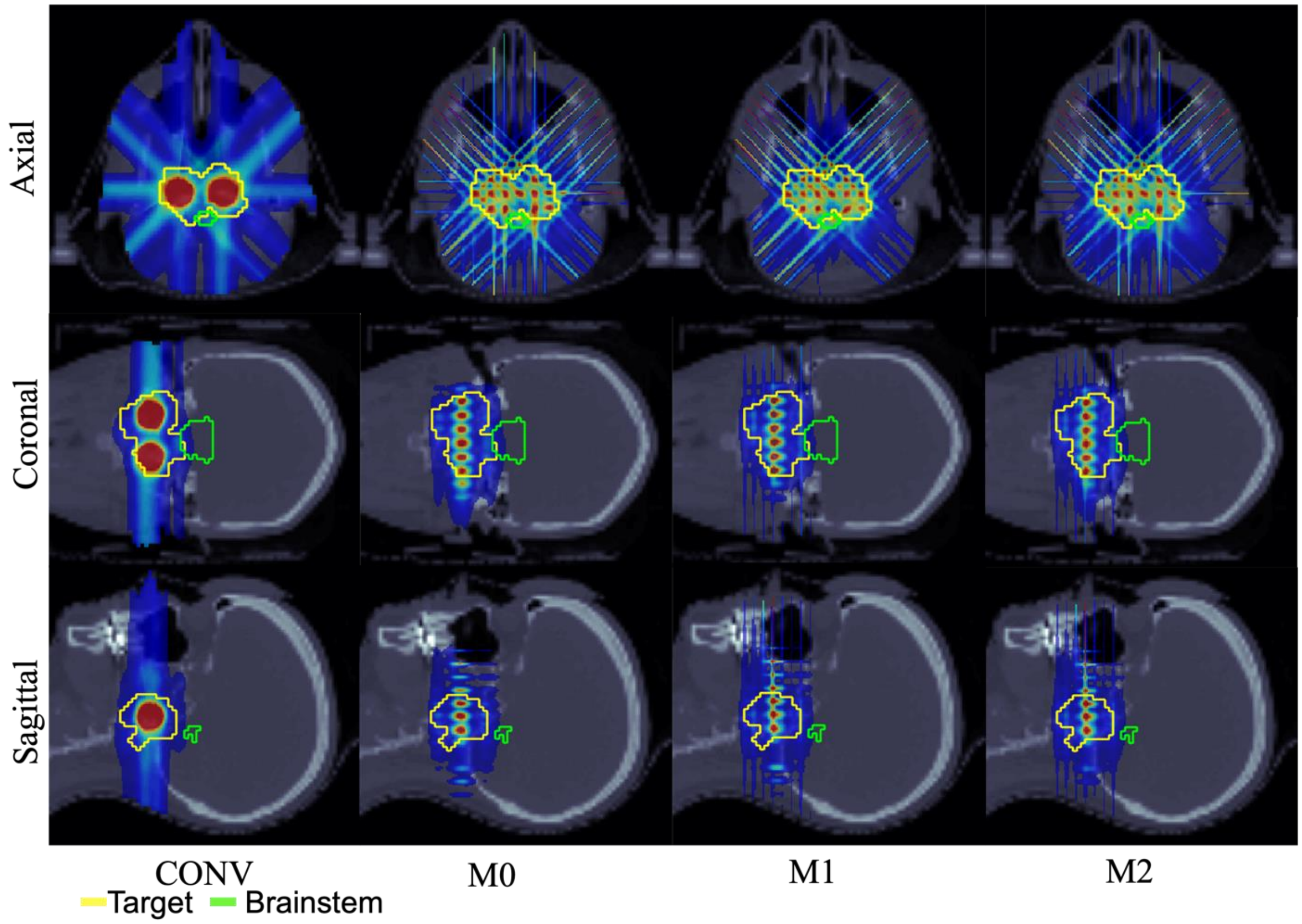

Figure 2. **HN01**. Dose plots: conventional pLATTICE (CONV) and minibeam-pLATTICE (M0, M1, M2).

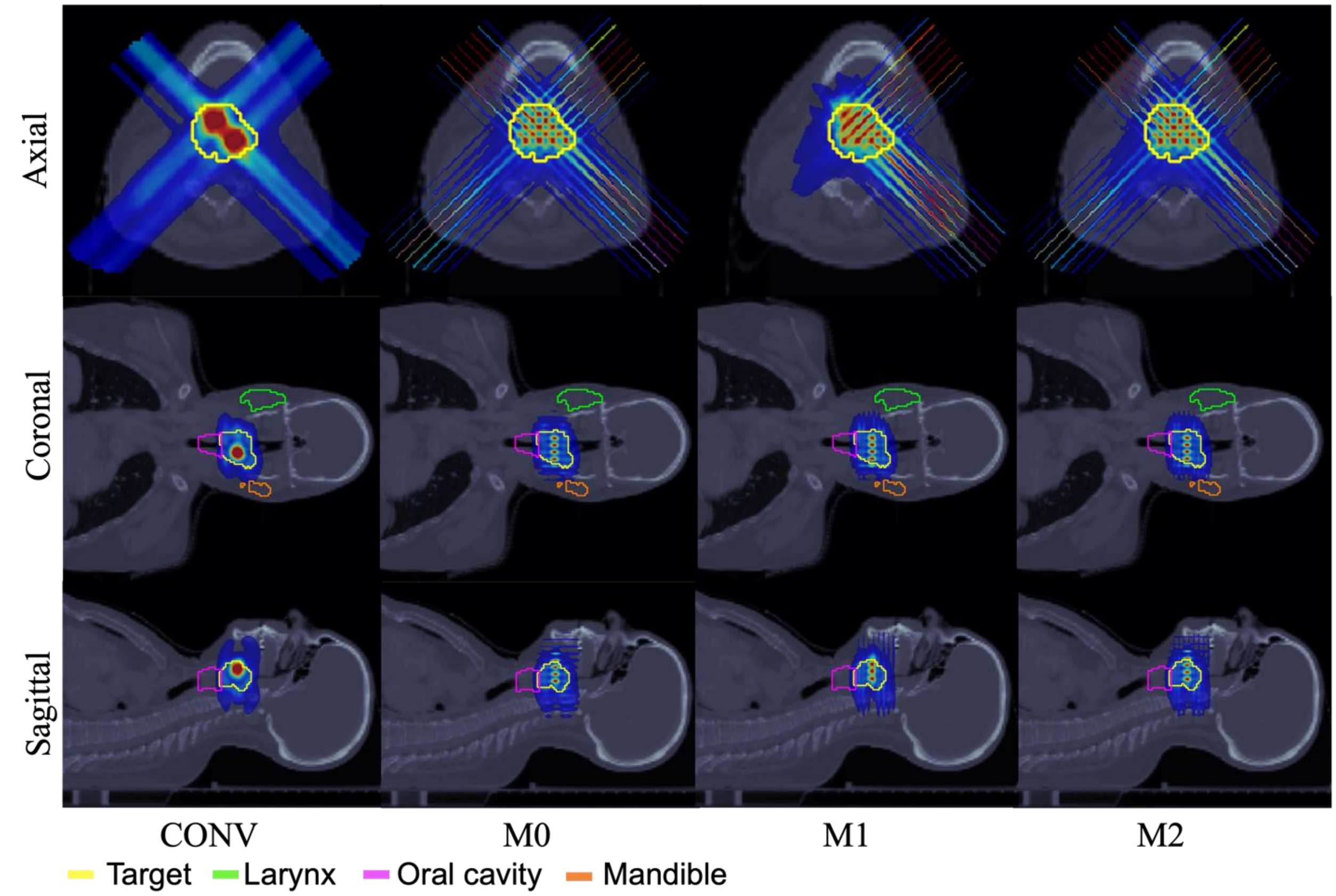

Figure 3. **HN02**. Dose plots: conventional pLATTICE (CONV) and minibeam-pLATTICE (M0, M1, M2).

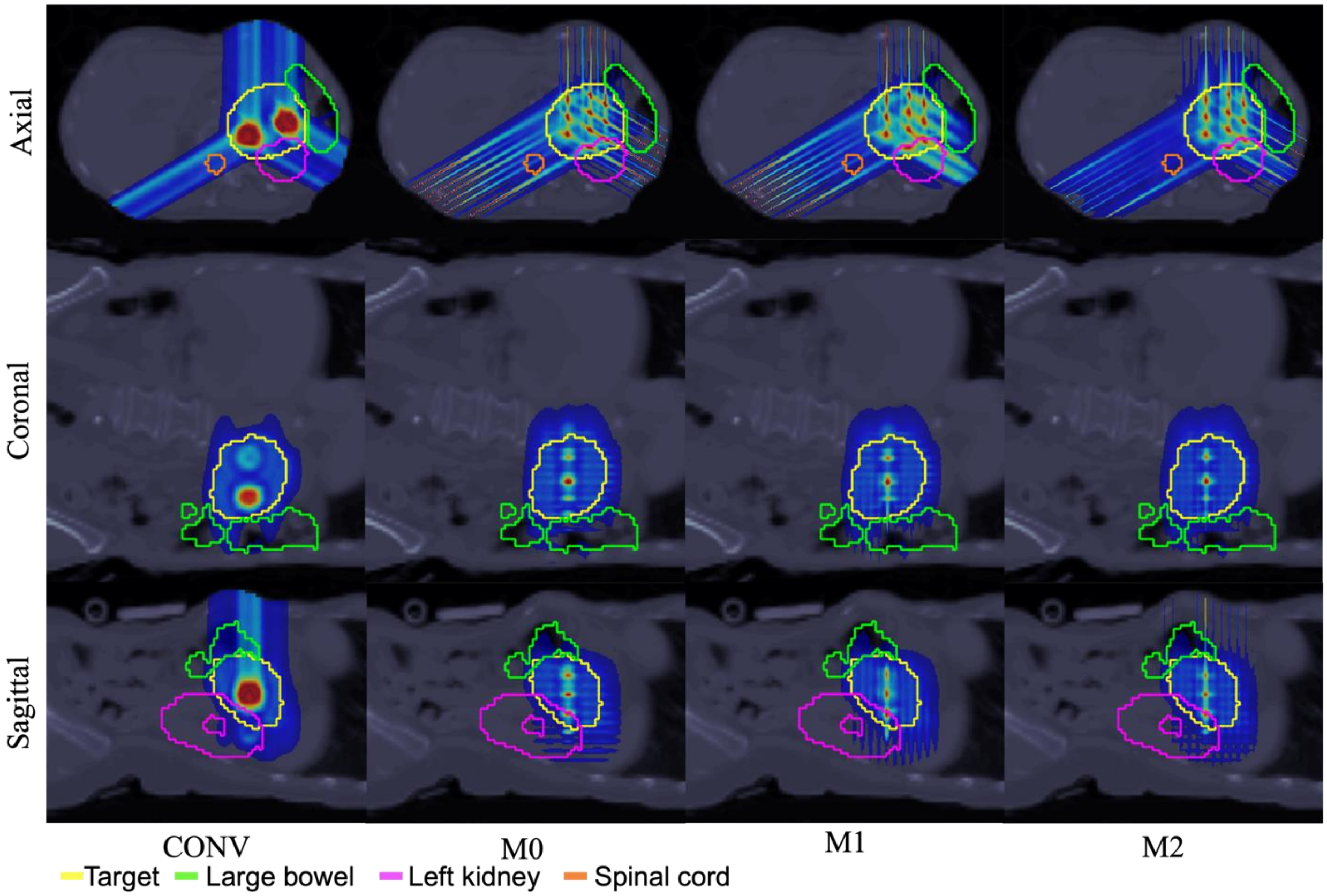

Figure 4: **Abdomen.** Dose plots: conventional pLATTICE (CONV) and minibeam-pLATTICE (M0, M1, M2).

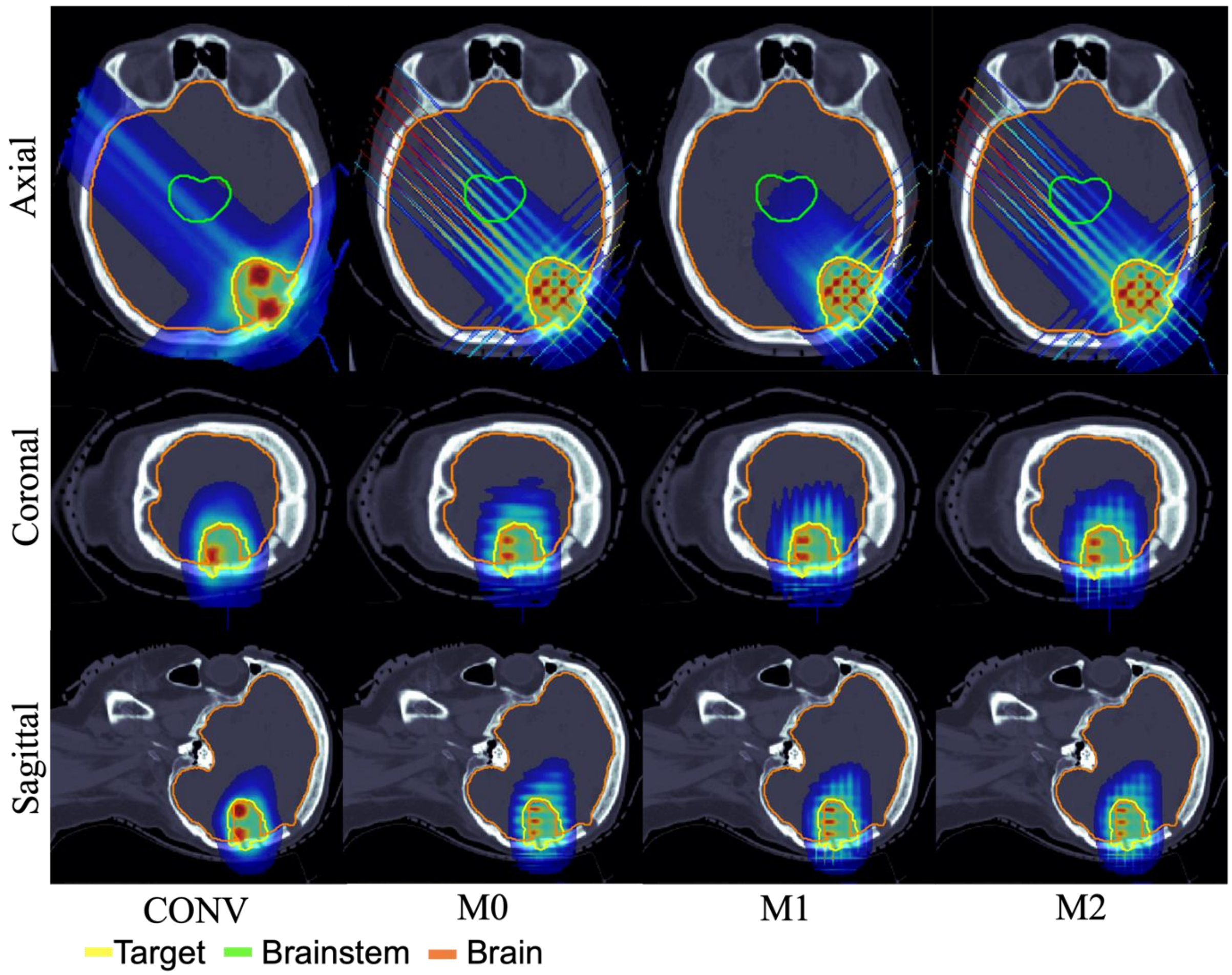

Figure 5: **Brain.** Dose plots: conventional pLATTICE (CONV) and minibeam-pLATTICE (M0, M1, M2).

**4. Discussion**

This study introduces a novel approach that integrates pMBRT with LATTICE radiation therapy (LRT) for treating small-to-medium-sized tumors. The framework combines the strengths of both modalities by optimizing spatial dose separation and target coverage through inverse planning tailored for proton minibeams. The optimization model and inverse planning strategy is particularly advantageous for models like M2, which require simultaneous dose modulation to improve peak and valley dose shaping and OAR sparing across multiple beam orientations. However, several important challenges and limitations remain, both in terms of biological validation and clinical feasibility.

Maintaining a high PVDR is crucial in LRT's intended radiobiological effectiveness. A high PVDR enhances tumor control while reducing toxicity to surrounding normal tissues due to the bystander and abscopal effects [36 37]. Preclinical studies have further demonstrated that high-PVDR minibeams also enhance immune response via increased calreticulin exposure, γH2AX signaling, and CD8+ T-cell infiltration [38, 39]. Recent studies [32, 40, 41] propose a joint treatment planning approach that optimizes dose conformity and dose-volume histogram (DVH)-based constraints while maximizing PVDR. Their findings demonstrate that this joint optimization improves target dose uniformity while achieving high PVDR at organs at risk (OAR). Additionally, [32] introduces a multi-collimator pMBRT approach that employs a set of multi-slit collimators (MSC) with varying center-to-center (ctc) distances at each beam angle, offering provable advantages over single-collimator methods. Integrating such PVDR optimization techniques with the minibeam-pLATTICE modality could further enhance PVDR in OAR while maintaining or improving target dose distribution.

While the results from the four anatomically distinct clinical test cases demonstrate the feasibility of minibeam-pLATTICE planning in complex regions, these cases might not be sufficient to establish statistical significance or assess interpatient anatomical variability. Differences in tumor location, size, and proximity to OAR across patients necessitate broader validation across a larger set of clinical test cases to evaluate robustness, enable formal statistical comparisons, and quantify variability in PVDR and OAR sparing. Future work will focus on expanding the study to a representative clinical dataset to support power analysis and further generalize the findings.

From a delivery perspective, the M2 model's requirement for dual collimator orientations per gantry angle increases treatment setup time, and the number of distinct QA configurations. These added demands may increase the QA workload and challenge existing hardware. Evaluating the total beam-on time, and setup efficiency will be necessary to determine clinical viability and maintain short treatment session length to reduce patient discomfort. Furthermore, successful delivery of minibeam-pLATTICE plans, particularly in the M2 model, requires sub-millimeter alignment and accurate orientation switching of MSC across gantry angles. Recent work [42] demonstrated the mechanical feasibility and reproducibility of MSC alignment across gantry rotations in a single-gantry proton facility. Using film-based dosimetric measurements, [42] work reported slit positioning reproducibility with consistent performance within 5%. These results indicate that precise multi-orientation collimator delivery is technically achievable and that the mechanical tolerances required by our M2 model are within reach of current system capabilities, though further hardware development and validation would be necessary for clinical integration.

Furthermore, while this study validates the feasibility of minibeam-pLATTICE through computational modeling, experimental validation [42-44] is necessary to facilitate clinical translation. Recently, [42] has shown that minibeam delivery systems, can be successfully constructed and commissioned on clinical proton gantries. Complementary works [43, 44] employing film-based dosimetry in water-equivalent phantoms have validated key physical characteristics, such as PVDR preservation, and slit alignment reproducibility, across multiple gantry angles. Together, these studies provide critical groundwork for bridging treatment planning with experimental implementation and ensuring micro-dosimetric accuracy, which are not fully captured by the MatRad pencil beam dose engine.

Another critical challenge in pencil beam scanning proton therapy is precise spot placement, especially in LATTICE RT, where sharp dose gradients at peak boundaries are essential. Recent works [45-47] have proposed promising strategies for adaptive spot placement and mixed spot sizes to improve dose conformality and reduce valley fill-in. For example, gradient-based or multiscale spot allocation algorithms [45, 46] can increase spot density in high-gradient regions, while dual-kernel strategies [47] offer coarse valley coverage and fine peak modulation. However, most adaptive spot placement techniques are implemented in 2D or for homogenous dose distributions, whereas the lattice peaks in our model require coordinated 3D optimization across multiple beam angles and orientations. Implementing dual-kernel spot sizing also requires customized dose kernel libraries and delivery sequencing. While these approaches are technically innovative and conceptually align with the goal of enhancing spatial modulation, their integration into our current 3D inverse planning framework presents significant challenges. Future work may explore incorporating these techniques to further refine dose gradients and delivery efficiency in minibeam-pLATTICE planning.

Overall, this work demonstrates the promise of extending LRT to small-to-medium target volumes using proton minibeams. The proposed minibeam-pLATTICE framework not only achieves favorable dosimetric outcomes but also introduces a novel optimization paradigm that enables fine-grained spatial dose modulation through inverse planning. While we acknowledge that experimental validations, and biological modeling remain as future work, the computational findings presented here establish a compelling proof of concept. The methodology provides a strong foundation for subsequent translational studies and offers new insights into how spatially fractionated proton therapy can be adapted to anatomically constrained clinical scenarios.

## 5. Conclusion

This study aimed to extend the LATTICE radiation therapy paradigm to treat small-to-medium-sized target volumes. To achieve this, a novel proton LATTICE approach using minibeams, minibeam-pLATTICE, was introduced. Three models were developed, each incorporating different collimator orientations relative to the proton beam source. Using inverse planning, the approach successfully generated dose plans with distinct lattice peak and valley regions for clinical head-and-neck cases with small target volumes. To the best of our knowledge, this is the first study to apply the LATTICE therapy to small-to-medium-sized tumors using proton minibeams. While this work is primarily computational, it establishes a foundation for future translational efforts by highlighting the optimization and delivery trade-offs unique to the minibeam-pLATTICE setting.

**Acknowledgments**

The authors are very thankful for the valuable comments from reviewers. This research is partially supported by the NIH grants No. R37CA250921, R01CA261964, and a KUCC physicist-scientist recruiting grant.

**Conflict of Interest Statement**

None.